\documentclass[11pt]{article}
\usepackage{graphicx}
\usepackage{amsmath}
\usepackage{amsfonts}
\usepackage{amssymb}
\usepackage{epsfig}
\usepackage{times}
\usepackage{cite}
\usepackage{calc}
\usepackage{version}
\usepackage[french,english]{babel}
\newcommand{\lqcd}{\Lambda_{_{_{\rm QCD}}}}

\begin{document}


\begin{titlepage}


\null

\vskip 2.5cm

{\bf\large\baselineskip 20pt
\begin{center}
\begin{Large}
Medium-modified average multiplicity and multiplicity fluctuations in jets
\end{Large}
\end{center}
}
\vskip 1cm

\begin{center}

Redamy P\'erez-Ramos\footnote{E-mail: redamy@mail.desy.de}\\
\smallskip
II. Institut f\"ur Theoretische Physik, Universit\"at Hamburg\\
Luruper Chaussee 149, D-22761 Hamburg, Germany
\end{center}

\baselineskip=15pt

\vskip 3.5cm

{\bf Abstract}: The energy evolution of average multiplicities 
and multiplicity fluctuations in 
jets produced in heavy-ion collisions is investigated from 
a toy QCD-inspired model. 
In this model, we use modified splitting functions 
accounting for medium-enhanced radiation of gluons 
by a fast parton which propagates through the
quark gluon plasma. The leading contribution of the standard production of soft
hadrons is enhanced by a factor $\sqrt{N_s}$ while
next-to-leading order (NLO) corrections are suppressed by $1/\sqrt{N_s}$, where
the parameter $N_s>1$ accounts for the 
induced-soft gluons in the medium.
Our results for such global observables are 
cross-checked and compared 
with their limits in the vacuum.

\end{titlepage}

Recent experiments at the Relativistic Heavy Ion Collider (RHIC) have established
a phenomenon of strong high-transverse momentum hadron suppression 
\cite{PHENIX}, which supports the picture 
that hard partons going through dense matter
suffer a significant energy loss prior to hadronization
in the vacuum (for recent review see \cite{arleo}).

Predictions concerning multi-particle production
in nucleus-nucleus collisions can be carried out by using a toy QCD-inspired model 
introduced by Borghini and Wiedemann in \cite{Urs};
it allows for analytical computations and may
capture some important features of a more complete QCD description. 
In this model, the Dokshitzer-Gribov-Lipatov-Altarelli-Parisi 
(DGLAP) splitting functions $q\to g\bar q$ and $g\to gg$ \cite{Basics}
of the QCD evolution equations 
were distorted so that the role of soft emissions was
enhanced by multiplying the infra-red
singular terms by the medium factor $N_s$. 
The model \cite{Urs} was further discussed and used on the 
description of final states hadrons produced in heavy-ion 
collisions \cite{sapeta}. 

Within the model, we make predictions for the medium-modified 
average multiplicity $N_A$ in quark and gluon jets ($A=q,g$)
produced in such reactions, 
for the ratio $r=N_g/N_q$ 
and finally for the second multiplicity correlators 
$\langle N_A(N_A-1)\rangle/N_A^2$,
which determines the width of the multiplicity distribution.

The starting point of our analysis is the NLO or
Modified-Leading-Logarithmic-Approximation (MLLA)
master evolution equation for the
generating functional \cite{Basics}
which determine the jet properties
at all energies together with the initial conditions at 
threshold at small $x$, where $x$ is the fraction of the 
outgoing jet energy carried away by a single gluon.
Their solutions with medium-modified splitting functions
can be resummed in powers of
$\sqrt{\alpha_s/N_s}$ and the leading contribution can be 
represented as an exponential of the medium-modified anomalous 
dimension which takes into account the $N_s$-dependence:
\begin{equation}\label{eq:intrep}
N_A\simeq \exp\left\{\int^{Y}
\gamma_{\text{med}}\left(\alpha_s(Y)\right)dY\right\},
\end{equation}
where $\gamma_{\text{med}}(\alpha_s)$ can be expressed as
a power series of $\sqrt{\alpha_s/N_s}$
in the symbolic form:
$$
\gamma_{\text{med}}\left(\alpha_s\right)
\simeq\sqrt{N_s}\times\sqrt{\alpha_s}\left(1+
\sqrt{\frac{\alpha_s}{N_s}}+
{\cal O}\left(\frac{\alpha_s}{N_s}\right)\right).
$$
Within this logic, the leading double logarithmic approximation 
(DLA, ${\cal O}(\sqrt{N_s\alpha_s})$), which resums both
soft and collinear gluons, and NLO 
(MLLA, ${\cal O}(\alpha_s)$), which resums hard collinear partons
and accounts for the running of the coupling constant $\alpha_s$,
are complete.  
The choice $dY=d\Theta/\Theta$, 
where $\Theta\ll1$ is the angle between outgoing
couples of partons in independent partonic emissions, follows from Angular
Ordering (AO) in intra-jet cascades \cite{Basics}.
In order to obtain the hadronic spectra,
we advocate for the Local Parton Hadron Duality (LPHD) hypothesis \cite{LPHD}:
global and differential partonic observables can be normalized 
to the corresponding hadronic observables via a certain constant 
${\cal K}$ that can be fitted to the data, i.e. $N_{g,q}^h={\cal K}\times N_{g,q}$.

The evolution of a jet of energy $E$ and half-opening angle
$\Theta$ involves the DLA anomalous dimension $\gamma_0$ related
to the coupling constant $\alpha_s$ through $\gamma_0^2=2N_c\alpha_s/\pi$,
with $\alpha_s=2\pi/4N_c\beta_0(Y+\lambda)$, where $Y=\ln(Q/Q_0)$ 
($Q=E\Theta$ is the hardness or maximum transverse momentum of the jet),
$\lambda=\ln(Q_0/\lqcd)$ is a parameter associated with 
hadronization ($Q_0$ is the collinear cut-off parameter,
$k_T>Q_0$, and $\lqcd$ is the intrinsic QCD scale) and
$\beta_0=\frac1{4N_c}\left(\frac{11}{3}N_c-\frac43T_R\right)$,
where $T_R=n_f/2$, $n_f$ being the number of active flavors.
At MLLA, as a consequence of angular ordering in parton cascading, the average
multiplicity inside a gluon and a quark jet, $N_{g,q}$, obey the system
of two-coupled evolution equations \cite{DREMIN} (the subscript $_Y$ 
denotes $d/d Y$)
\begin{eqnarray}
{N_g}_Y\!\!&\!\!=\!\!&\!
\!\int_0^1 dx\,\gamma_0^2\left[\Phi_g^g\left(N_g(x)+
N_g(1-x)-N_g\right)+n_f\Phi_g^q
\left({N_q}(x)+
N_q(1-x)-N_g\right)\right],\label{eq:NGh}\\
{N_q}_Y\!\!&\!\!=\!\!&\!\!\int_0^1 dx\,
\gamma_0^2\!\!\left[\Phi_q^g\left(N_g(x)+
N_q(1-x)-N_g\right)\right],
\label{eq:NQh}
\end{eqnarray}
which follow from the MLLA master evolution equation
for the generating functional; $N_{g,q}\equiv N_{g,q}(Y)$, 
$N_{g,q}(x)\equiv N_{g,q}(Y+\ln x)$,
$N_{g,q}(1-x)\equiv N_{g,q}(Y+\ln(1-x))$, 
$\Phi_A^B$ denotes the medium-modified DGLAP splitting
functions:
\begin{eqnarray}
\Phi_g^g(x)\!&\!=\!&\!\frac{N_s}x-(1-x)[2-x(1-x)],\notag\\
\Phi_g^q(x)\!&\!=\!&\!\frac1{4N_c}[x^2+(1-x)^2],\notag\\
\Phi_q^g(x)\!&\!=\!&\!\frac{C_F}{N_c}
\left(\frac{N_s}x-1+\frac{x}2\right),\label{eq:splitfunct}
\end{eqnarray}
which accounts 
for parton energy loss in the medium by enhancing the singular terms
like $\Phi\approx N_s/x$ as $x\ll1$
as proposed in the Borghini-Wiedemann 
model \cite{Urs}. Thus, when $N_s$ increases the 
DLA becomes dominant and energy-momentum conservation plays 
a less important role.

For $Y\gg\ln x\sim\ln(1-x)$, 
$N_{g,q}(x)$ ($N_{g,q}(1-x)$) can be
replaced by $N_{g,q}$ in the hard partonic splitting region
$x\sim1-x\sim1$ (non-singular or regular parts of the splitting functions), 
while  the dependence at small $x\ll1$ is kept in the singular
term $\Phi(x)\approx N_s/x$ as done in the vacuum. Furthermore, the integration
over $x$ can be replaced by the integration over 
$Y(x)=\ln\left(\frac{xE\Theta}{Q_0}\right)$. Thus, 
one is left with the approximate system of two-coupled equations,
\begin{eqnarray}\label{eq:NGhbis}
\frac{d^2}{dY^{2}}N_g(Y)\!&\!=\!&\!
\gamma_0^2\left(N_s-a_1\frac{d}{dY}\right)N_g(Y),\\
\frac{d^2}{dY^{2}}N_q(Y)\!&\!=\!&\!\frac{C_F}{N_c}\gamma_0^2
\left(N_s-\tilde a_1\frac{d}{dY}\right)N_g(Y),\label{eq:NQhbis}
\end{eqnarray}
with the initial conditions at threshold 
$N_A(0)=1$ and $N_A^{'}(0)=0$
and the hard constants: 
$$
a_1\!\!=\!\!\frac{1}{4N_c}\left[\frac{11}{3}N_c+
\frac43T_R\left(1-2\frac{C_F}{N_c}\right)\right],\quad  \tilde a_1=3/4.
$$ 
The quantum corrections $\propto a_1,\;\tilde a_1$
in (\ref{eq:NGhbis},\ref{eq:NQhbis}) arise from the integration 
over the regular part of the splitting functions, they are 
${\cal O}(\sqrt{\alpha_s/N_s})$ suppressed and {\em partially} account
for energy conservation as happens 
in the vacuum. 

\begin{figure}[t]
\begin{center}
\includegraphics[height=5cm]{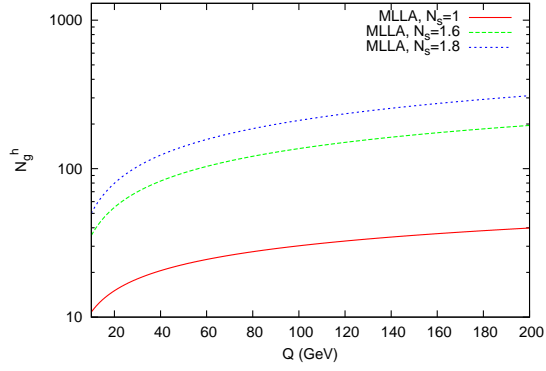}
\caption{MLLA (\ref{eq:nsmultg}) 
medium-modified average multiplicity as a function
of $Q=E\Theta$ in the vacuum ($N_s=1$) and
in the medium ($N_s=1.6$ and $N_s=1.8$) for $n_f=3$.
\label{fig:avemult}
}
\end{center}
\end{figure}

These equations can be solved
by applying the inverse Mellin transform
to the self-contained gluonic equation (\ref{eq:NGhbis}), which leads to
\begin{equation}\label{eq:intrepres}
N_g^h(Y)\simeq{\cal K}\times\int_C\frac{d\omega}{2\pi i}\omega^{\frac{a_1}{\beta_0}-2}
\exp{\left[\omega(Y+\lambda)+\frac{N_s}{\beta_0\omega}\right]},
\end{equation}
where the contour $C$ lies to the 
right of all singularities of $N_g(\omega)$
in the complex plane. 
Since we are concerned with
the asymptotic solution of the equation
as $Y\gg1$ ($E\Theta\gg Q_0$), that is the high-energy limit,
the inverse Mellin transform (\ref{eq:intrepres}) can be estimated
by the steepest descent method. Indeed, the large parameter is $Y$ and
the function in the exponent
presents a saddle point at 
$\omega_0=\sqrt{N_s/\beta_0(Y+\lambda)}$, such that 
the asymptotic solution reads
\begin{equation}\label{eq:nsmultg}
N_g^h(Y)\simeq {\cal K}\times 
{(Y+\lambda)}^{-\frac{\sigma_1}{\beta_0}}
\exp{\sqrt{\frac{4N_s}{\beta_0}(Y+\lambda)}},
\end{equation}
where $\sigma_1=\frac{a_1}2-\frac{\beta_0}4$.
The constant $\sigma_1$ is $N_s$-independent 
because it resums vacuum corrections. 
Therefore, the production of soft gluons in the medium becomes
$\exp\left[2(\sqrt{N_s}-1)\sqrt{(Y+\lambda)/\beta_0}\right]$ higher than the
standard production of soft gluons in the vacuum \cite{Basics}. 
From (\ref{eq:intrep}) 
and (\ref{eq:nsmultg}) one obtains the medium-modified MLLA
anomalous dimension
$\gamma_{\text{med}}=\frac{1}{N_g}\frac{d N_g}{d Y}=
\sqrt{N_s}\gamma_0-\sigma_1\gamma_0^2$, which
 is nothing but the MLLA rate
of multi-particle production with respect to the {\em evolution-time}
variable $Y$ in the dense medium.
In Fig.\,\ref{fig:avemult}, 
we display the medium-modified 
average multiplicity (\ref{eq:nsmultg}) with predictions
in the vacuum ($N_s=1$)
in the range $10\leq Q(\text{GeV})\leq200$; we set
$Q_0=\lqcd=0.23$ GeV in the limiting spectrum
approximation \cite{DREMIN}, and take ${\cal K}=0.2$ 
from \cite{DREMIN}. The values $N_s=1.6$ and $N_s=1.8$
in the medium may be realistic for RHIC and LHC phenomenology \cite{Urs,sapeta};
the jet energy subrange $10\leq Q(\text{GeV})\leq50$ displayed in 
Fig.\,\ref{fig:avemult} has been recently considered by the STAR collaboration, which
reported the first measurements of charged hadrons and particle-identified
fragmentation functions from p+p collisions \cite{heinz} at $\sqrt{s_{\text{NN}}}=200$ GeV. Finally, the whole jet energy range in the same figure, 
in particular for those values at $Q\geq50$ GeV, will
be reached at the LHC, i.e $Q=100$ GeV is an 
accessible value in this experiment (see \cite{Urs}
and references therein).

\begin{figure}[t]
\begin{center}
\includegraphics[height=5cm]{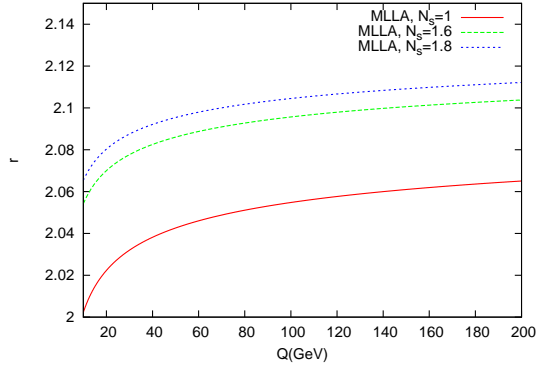}
\caption{MLLA ratio $r$ (\ref{eq:mllaratio}) 
as a function of $Q=E\Theta$ in the vacuum ($N_s=1$) and
in the medium ($N_s=1.6$ and $N_s=1.8$) for $n_f=3$.
\label{fig:ratio}
}
\end{center}
\end{figure}

We find, as expected, 
that the production of soft
hadrons increases as $N_s>1$: the available
phase space for the production of harder collinear hadrons
is restricted as the model itself states.
The medium-modified MLLA
gluon to quark average multiplicity ratio,
$r=N_g/N_q=N_g^h/N_q^h$, following from (\ref{eq:nsmultg}) and 
(\ref{eq:NQh}) reads
\begin{equation}\label{eq:mllaratio}
r=r_0\left[1-r_1\frac{\gamma_0}{\sqrt{N_s}}
+{\cal O}\left(\frac{\gamma_0^2}{N_s}\right)\right], \;
r_0=\frac{N_c}{C_F},
\end{equation}
where we introduced the coefficient
$r_1=a_1-\tilde a_1$ in the term suppressed by $\gamma_0/\sqrt{N_s}$ 
as $N_s>1$. Therefore, if compared with its behavior at $N_s=1$, we check, as expected
from the model \cite{Urs}, that $r$ becomes closer to its asymptotic DLA limit $r_0=N_c/C_F=9/4$, as depicted in Fig.\,\ref{fig:ratio}. 
Setting $N_s=1$ in (\ref{eq:mllaratio}), one recovers the appropriate limits
in the vacuum \cite{Basics,MALAZA,KhozeOchs}.
Finally, the gluon jets are still more active
than the quark jets in producing secondary particles  and the shape of the curves
are roughly the same.

The normalized second multiplicity
correlator $A_2=\langle {N_A}({N_A}-1)\rangle/N_A^2$ 
defines the width of the multiplicity distribution and is related to its 
dispersion by the formula $D_A^2=(A_2-1)N_A^2+N_A$ \cite{MALAZA}.
These moments, which are less inclusive than the average multiplicity, 
prove to be ${\cal K}$-independent and therefore provide a pure test of multiparticle
production. 
The medium-modified system of two-coupled evolution equations for this observable
follows from the MLLA master equation for the azimuthally averaged generating
functional \cite{Basics} and can be written in the convenient form

\vbox{
\begin{eqnarray}
\frac{d}{dY}(N_g^{(2)}-N_g^2)\!\!&\!\!=\!\!&\!\!\int_0^1dx\gamma_0^2\Phi_g^g\left[N_g^{(2)}
(Y+\ln x)\!+\!\Big(N_g^{(2)}(Y+\ln(1-x))-N_g^{(2)}(Y)\Big)\right.\cr
&&\left.\hskip 0.5cm+\Big(N_g(Y+\ln x)-N_g(Y)\Big)
\Big(N_g(Y+\ln(1-x))-N_g(Y)\Big)\right]\cr
&&\hskip -1cm+n_f\!\!\int_0^1dx\gamma_0^2\Phi_g^q
\left[2\Big(N_q^{(2)}(Y+\ln x)\!-\!N_q^2(Y+\ln x)\Big)
\!-\!\Big(N_g^{(2)}(Y)\!-\!
N_g^2(Y)\Big)\right.\cr
\!\!&\!\!+\!\!&\!\!\left.\Big(2N_q(Y+\ln x)-N_g(Y)\Big)
\Big(2N_q(Y+\ln(1-x))-N_g(Y)\Big)\right]\label{eq:corrg},\\
\frac{d}{dY}(N_q^{(2)}-N_q^2)\!\!&\!\!=\!\!&\!\!\int_0^1dx\gamma_0^2
\Phi_q^g
\left[N_g^{(2)}(Y+\ln x)\!+\!\left(N_q^{(2)}(Y+\ln(1-x))-N_q^{(2)}(Y)\right)\right.\cr
&&\left.\hskip 0.3cm+2\Big(N_g(Y+\ln x)- N_q(Y)\Big)
\Big(N_q(Y+\ln(1-x))- N_q(Y)\Big)\right]\label{eq:corrq},
\end{eqnarray}
}
which proves to be more suitable for obtaining analytical solutions
in the following.
We use a new method 
to compute solutions at MLLA by replacing
$N_A^{(2)}=A_2N_A^2$ on both sides of the expanded equations at $x\sim1-x\sim1$. 
The notations in 
(\ref{eq:corrg},\ref{eq:corrq}) follow the same logic as those
in (\ref{eq:NGhbis},\ref{eq:NQhbis}).
Applying the analysis that led to the system 
(\ref{eq:NGhbis},\ref{eq:NQhbis}),
we obtain from (\ref{eq:corrg},\ref{eq:corrq})
\begin{eqnarray}\label{eq:N2G}
\frac{d^2}{dY^2}\left(N^{(2)}_g-N_g^2\right)
\!&\!=\!&\!\gamma_0^2\left(N_s-a_1\frac{d}{dY}\right)N^{(2)}_g+
(a_1-b_1)\gamma_0^2\frac{d}{dY} N_g^2,\\
\frac{d^2}{dY^2}\left(N^{(2)}_q-N_q^2\right)\!&\!=\!&\!
\frac{C_F}{N_c}\gamma_0^2\left(N_s-
\tilde a_1\frac{d}{dY}\right)N^{(2)}_g,
\label{eq:N2Q}
\end{eqnarray}
where 
$$
b_1=\frac1{4N_c}\left[
\frac{11}{3}N_c-4\frac{T_R}{N_c}\left(1-2\frac{C_F}{N_c}\right)^2\right].
$$ 
The constant $N_s$ only affects the leading double logarithmic term of the
equations. The terms proportional to $a_1$, $(a_1-b_1)$ and $\tilde a_1$ are
hard vacuum corrections, which {\em partially} account for energy conservation, indeed
$\gamma_0^2\frac{dN}{dY}\approx\sqrt{N_s}\gamma_0^3$ and the relative correction
to DLA is ${\cal O}(\sqrt{\alpha_s/N_s})$. 

\begin{figure}[t]
\begin{center}
\includegraphics[height=5cm]{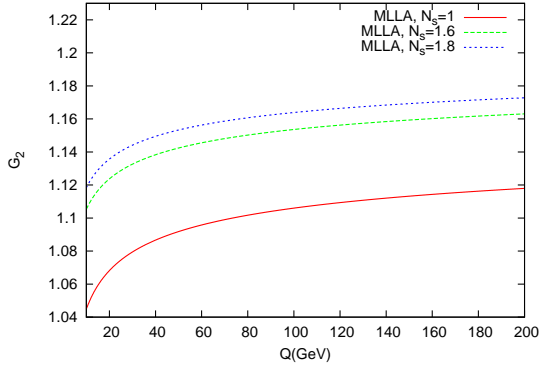}
\caption{MLLA 
second multiplicity correlator inside a gluon jet (\ref{eq:expG2}) 
as a function of $Q=E\Theta$ in the vacuum ($N_s=1$) and
in the medium ($N_s=1.6$ and $N_s=1.8$ for $n_f=3$.
\label{fig:G2}
}
\end{center}
\end{figure}

\begin{figure}[t]
\begin{center}
\includegraphics[height=5cm]{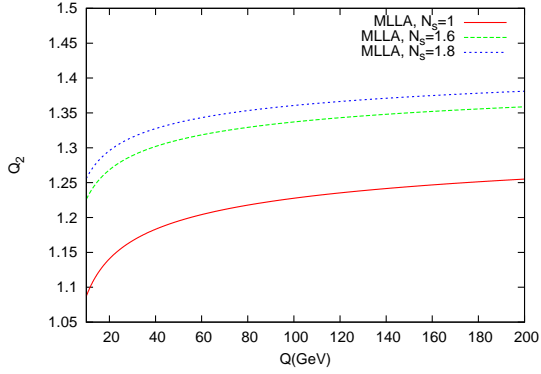}
\caption{MLLA 
second multiplicity correlator inside a quark jet (\ref{eq:expQ2}).
\label{fig:Q2}
}
\end{center}
\end{figure}

Setting $N^{(2)}_g=G_2N_g^2$ in (\ref{eq:N2G}) and
making use of (\ref{eq:nsmultg}), the system can be solved 
iteratively by taking terms up to ${\cal O}(\alpha_s)$ into
consideration. The analytical solution reads,
\begin{equation}\label{eq:G2MLLA}
G_2-1=
\frac{1-\displaystyle{\left(\frac23a_1+2b_1\right)\frac{\gamma_0}{\sqrt{N_s}}}
+{\cal O}\left(\frac{\gamma_0^2}{N_s}\right)}{3-(4a_1-\beta_0)\displaystyle{\frac{\gamma_0}{\sqrt{N_s}}}+{\cal O}\left(\frac{\gamma_0^2}{N_s}\right)},
\end{equation}
while its expansion in the form 
$1+\gamma_0/\sqrt{N_s}$ leads to
\begin{equation}\label{eq:expG2}
G_2-1\approx\frac13-c_1\frac{\gamma_0}{\sqrt{N_s}}+
{\cal O}\left(\frac{\gamma_0^2}{N_s}\right),
\end{equation}
where the linear combination of color factors can be written in the form
\begin{equation}\label{eq:c1}
c_1
=\frac{1}{4N_c}\!\left(\!\frac{55}{9}-4\frac{T_R}{N_c}
+\frac{112}{9}\frac{T_R}{N_c}\frac{C_F}{N_c}
-\frac{32}{3}\frac{T_R}{N_c}\frac{C_F^2}{N_c^2}\right).
\end{equation}
We use (\ref{eq:expG2}) and (\ref{eq:mllaratio}) and substitute
$N^{(2)}_q=Q_2N_q^2$ into (\ref{eq:N2Q}) such that the solution reads
\begin{eqnarray}
Q_2-1\approx\frac{N_c}{C_F}
\left(\frac1{3}-\tilde c_1\frac{\gamma_0}{\sqrt{N_s}}\right)
+{\cal O}\left(\frac{\gamma_0^2}{N_s}\right),
\label{eq:expQ2}
\end{eqnarray}
where
we obtain the combination of color factors
\begin{eqnarray}
\tilde c_1=\frac{1}{4N_c}\!
\left(\!\frac{55}{9}+\frac49\frac{T_R}{N_c}\frac{C_F}{N_c}
-\frac83\frac{T_R}{N_c}\frac{C_F^2}{N_c^2}\right).
\end{eqnarray}
Setting $N_s=1$ in (\ref{eq:expG2}) and (\ref{eq:expQ2})
we get a perfect agreement with the vacuum results \cite{MALAZA}.
In Fig.\,\ref{fig:G2} 
and Fig.\,\ref{fig:Q2}, we compare
our results for the medium-modified
second multiplicity correlators (\ref{eq:expG2}) and (\ref{eq:expQ2})
with predictions in the vacuum ($N_s=1$) \cite{MALAZA}
in the limiting spectrum approximation inside the typical
range $10\leq Q(\text{GeV})\leq200$ for RHIC and LHC phenomenology.
Similarly to the MLLA ratio $r(N_s)$, Eq. (\ref{eq:mllaratio}), 
the hard corrections ${\cal O}(\gamma_0)$ are suppressed 
by a factor $1/\sqrt{N_s}$. 
As expected from the model, we check that
these results approach their DLA limits 
when $N_s$ increases; 
moreover, the multiplicity
fluctuations of individual events must be larger for quark
jets as compared to gluon jets just like in the vacuum \cite{MALAZA}.
Another interesting feature of these observables concerns the shape of the curves.
They are roughly identical and prove not to depend on the medium parameter
$N_s$. Moreover, there exists evidence for a flattening of the slopes 
as the hardness of the jet $Q=E\Theta$ increases for
$N_s\geq1$ (vacuum and medium). This kind of scaling behavior 
is known as the Koba-Nielsen-Olsen (KNO) scaling \cite{DokKNO}: 
it was discovered by Polyakov in quantum 
field theory \cite{Poly} and experimentally 
confirmed by $e^+e^-$ 
measurements \cite{experN2} for the second and higher order
multiplicity correlators.

In this paper we have dealt with the medium-modified average multiplicity 
and the medium-modified second multiplicity correlator in quark and gluon jets 
at RHIC and LHC energy scales. 
The starting point of our calculations is based on the 
Borghini-Wiedemann work
\cite{Urs}, which models parton energy loss 
in a nuclear medium.
The average multiplicity is found to be enhanced by the
factor $\sqrt{N_s}$ acting on the
exponential leading contribution (\ref{eq:nsmultg});
this leads in particular to the rescaling of the anomalous dimension $\gamma_{\text{med}}$ 
($\gamma\to\gamma_{\text{med}}\approx\sqrt{N_s}\gamma_0$) or equivalently,
to the enhancement of the in medium coupling constant.
Since hard corrections are suppressed
by the extra factor $1/\sqrt{N_s}$, 
it is straightforward to check that
$r$, $G_2$ and $Q_2$ approach the asymptotic DLA limits $r_0=N_c/C_F$,
$G_2=4/3$ and $Q_2=1+N_c/3C_F$ \cite{Basics} when $N_s$ increases. 
The previously mentioned KNO-scaling experienced by $G_2$ and 
$Q_2$ proves no special sensibility to the model and should normally hold like 
in the vacuum. 

Finally, since these results are model-dependent, they may still be improved
in the future, specially after the $N_s$-dependence 
of the non-singular parts 
of the splitting functions (\ref{eq:splitfunct}) 
has been exactly computed.

{\em Perspective:} 
Many experimental characterizations of the medium-modified intrajet
structure in heavy-ion collisions at RHIC and at the LHC require a soft
momentum cut-off $p_T^{\text{cut}}$, with $Q>p_T^{\text{cut}}$ to remove
the effects of the high multiplicity background. In \cite{Urs}, the soft background
was subtracted by integrating the single inclusive
differential distribution $\frac{dN}{d\ln p_T}$ (``hump-backed plateau")
over the range $Q\geq p_T\geq p_T^{\text{cut}}$,
with $p_T^{\text{cut}}>\Lambda_{QCD}$. Accordingly, the equivalent computation should
be performed for the second multiplicity correlator by integrating the double
differential inclusive distribution (two-particle correlation) 
$\frac{d^2N}{d\ln p_{1,T}d\ln p_{2,T}}$ over $p_{i,T}$, with the lower bounds
of integration $p_{i,T}^{\text{cut}}>\Lambda_{QCD}$ ($i=1,2$). 
Imposing such a cut-off in our calculations will affect the 
normalization rather than the behavior
and the shape of these observables as a function of $N_s$ and the
jet energy scale of the process $Q$ \cite{BRPR}.

%

\begin{thebibliography}{20}

\bibitem{PHENIX}
K. Adcox et al. (PHENIX Collab.), Phys. Rev. Lett. {\bf 88}
(2002) 022301;\newline
S.S. Adler et al. (PHENIX Collab.),
Phys. Rev. Lett. {\bf 91} (2003) 072301;\\
C. Adler et al. (STAR Collab.),
Phys. Rev. Lett {\bf 89} (2002) 202301.

\bibitem{arleo}
F. Arleo, hep-ph/0810.1193 and references therein;\\
S. Peign\'e \& A.V. Smilga, hep-ph/0810.5702.

\bibitem{Urs}
N. Borghini \& U.A. Wiedemann, hep-ph/0506218.

\bibitem{Basics}
Yu.L. Dokshitzer, V.A. Khoze, A.H. Mueller \& S.I. Troyan, Basics of
Perturbative QCD, Editions Fronti\`eres, Paris (1991).

\bibitem{sapeta}
S.Sapeta \& U.A Wiedemann, Eur. Phys. J. {\bf C55} (2008) 293.

\bibitem{LPHD}
Ya.I. Azimov, Yu.L. Dokshitzer, V.A. Khoze \& S.I. Troian,
Z. Phys. {\bf C} 27 (1985) 65;
Yu.L. Dokshitzer, V.A. Khoze \& S.I. Troian,
J. Phys. {\bf G} 17 (1991) 1585.

\bibitem{DREMIN}
I.M. Dremin, V.A. Nechitailo, Mod. Phys. Lett. A {\bf 9} (1994) 1471;
JETP Lett. {\bf 58} (1993) 945.

\bibitem{heinz}
M. Heinz for the STAR Collaboration, nucl-exp/0809.3769.

\bibitem{MALAZA}
E.D. Malaza \& B.R. Webber, Phys. Lett. {\bf B} 149 (1984) 501;
E.D. Malaza \& B.R. Webber, Nucl. Phys. {\bf B} 267 (1986) 702.

\bibitem{KhozeOchs}
V.A. Khoze \& W. Ochs,
Int. J. Mod. Phys. A {\bf 12} (1997) 2949.

\bibitem{DokKNO}
Yu.L. Dokshitzer, Phys.
Lett. B {\bf 305} (1993) 295.

\bibitem{Poly}
A.M. Polyakov, Sov. Phys. JETP {\bf 32} (1971) 296.

\bibitem{experN2}
HRS Coll., Phys. Rev. D {\bf 34} (1986) 3304;
AMY Coll., Phys. Rev. D {\bf 42} (1990) 737;
DELPHI Coll., Z. Phys. C-Particles and Fields
{\bf 50} (1991) 185.

\bibitem{BRPR}
N. Borghini \& R. Perez-Ramos, in preparation.



\end{thebibliography}
\end{document}